\documentclass[review]{elsarticle}
\usepackage{float}
\usepackage{geometry}
\usepackage{setspace}

\usepackage{epstopdf}
\usepackage{diagbox}
\usepackage{hyperref}
\usepackage{amsmath}

\journal{Journal of \LaTeX\ Templates}

\begin{document}

\begin{frontmatter}

\title{Quantifying time irreversibility using probabilistic differences between symmetric permutations}

\author[mymainaddress]{Wenpo Yao}
\author[myfirstaddress,mysecondaddress]{Wenli Yao}
\author[mythirdaddress]{Jun Wang\corref{mycorrespondingauthor}}
\cortext[mycorrespondingauthor]{Corresponding author}
\ead{wangj@njupt.edu.cn}
\author[myforthaddress]{Jiafei Dai}

\address[mymainaddress]{School of Telecommunications and Information Engineering, Nanjing University of Posts and Telecommunications, Nanjing 210003, Jiangsu, China}
\address[myfirstaddress]{School of Mines, China University of Mining and Technology, Xuzhou 221116, China}
\address[mysecondaddress]{Department of Mining and Metallurgical Engineering, Western Australian School Mines, Curtin University, Kalgoorlie, WA, Australia}
\address[mythirdaddress]{School of Geography and Biological Information, Nanjing University of Posts and Telecommunications, Nanjing 210023, Jiangsu, China}
\address[myforthaddress]{Nanjing General Hospital of Nanjing Military Command, Nanjing 210002, China}

\begin{abstract}
To simplify the quantification of time irreversibility, we employ order patterns instead of the raw multi-dimension vectors in time series, and considering the existence of forbidden permutation, we propose a subtraction-based parameter, $Y_{S}$, to measure the probabilistic differences between symmetric permutations for time irreversibility. Two chaotic models, the logistic and Henon systems, and reversible Gaussian process and their surrogate data are used to validate the time-irreversible measure, and time irreversibility of epileptic EEGs from Nanjing General Hospital is detected by the parameter. Test results prove that it is promising to quantify time irreversibility by measuring the subtraction-based probabilistic differences between symmetric order patterns, and our findings highlight the manifestation of nonlinearity of whether healthy or diseased EEGs and suggest that the epilepsy leads to a decline in the nonlinearity of brain electrical activities during seize-free intervals.
\end{abstract}


\end{frontmatter}

\section{Introduction}
Nonlinearity is a necessary condition for chaotic behaviors, so whether a time series is compatible with a linear process is important in the paradigm of deterministic chaos. Time irreversibility is one of fundamental properties that determine whether a time series is linear or not. Complex systems characterized with nonlinearity and nonequilibrium yield statistical asymmetry under time reversal, therefore, measurements for time irreversibility or directional asymmetry are feasible assessments for characterizing non-linear complex activities.

A stochastic process is said to be time reversible if its probabilistic properties are invariant with respect to time reversal, other-wise it is directional or irreversible \cite{Weiss1975,Kelly1979,Ramsey1995}. Recent reports describe the time irreversibility from different perspectives such as temporal asymmetry \cite{Graff2013,Costa2005}, complex network \cite{Donges2013}, visibility graph \cite{Lacasa2012,Flanagan2016}, entropy production \cite{Roldan2010,Roldan2012}, symbolic methods \cite{Kennel2004,Daw2000}, and others. Based on the theoretical definitions \cite{Ramsey1995}, measures for time irreversibility generally target on the divergences in joint probability distributions between the forward and backward time series \cite{Roldan2010,Hou2013,Jucha2014} or the probabilistic differences between symmetric distributions \cite{Costa2005,Costa2008,Guzik2006,Porta2008}. However, the estimation of joint probability from the raw time series (i.e., experimental signals) is not trivial. For the simplification of measuring time irreversibility, some measures based on symbolic dynamics \cite{Graff2013,Kennel2004,Daw2000,Hou2013} are proposed and have been gaining popularity for the features of simplicity, computationally efficiency, robustness, reduced demands on time series, and so no. The symbolic approaches provide a rigorous way of dealing with dynamics with finite precision rather than dig into raw time series \cite{Daw2003}, and in this present paper, we employ order patterns instead of vectors in raw the time series to simplify the quantification of time irreversibility.

For symbolic measures, it is not realizable to always have all of the designed symbolic sequences, and the nonoccurring sequences, called forbidden sequences or forbidden words, are closely related to the nonlinear complexity of the dynamical processes \cite{Daw2003}. In characterizing dynamical information by order patterns, the properties of forbidden order patterns have been given in-depth study and attention due to its close connections to dynamical complexity and structural information \cite{Amigo2007,Amigo2010,Carpi2010,Amigo2015,Kulp2017}. Considering the existence of forbidden words, practically speaking, vectors with some kinds of distributions may not have symmetric forms or may not have their corresponding forms in the reverse time series. Case like this makes the division-based measures, like relative entropy, Chernoff distance or their modified forms \cite{Johnson2001}, unsuitable for measuring the probabilistic difference between symbolic sequences while measurements based on subtraction should be reliable choices.

In our contributions, we propose a novel subtraction-based parameter to quantify temporal asymmetry by measuring the probabilistic differences between symmetric order patterns, and we verify the parameter by chaotic and reversible processes and apply it to characterize the nonlinearity of epileptic EEGs.

\section{Methodology}

\subsection{Statistical definitions of time irreversibility}
Let us recall some basic theoretical concepts of time irreversibility. In the statistical definition of G. Weiss\cite{Weiss1975}, if a process $X(t)$ is time reversal, vectors $\{X(t_{1}),X(t_{2}),\cdots,X(t_{m})\}$ and $\{X(-t_{1}),X(-t_{2}),\cdots,X(-t_{m})\}$ for every $t$ and $m$ have same joint probability distributions, say, it is invariant under the time-scale reversal.

Another definition of F. Kelly \cite{Kelly1979} suggests that a time series $X(t)$ is time reversible if $ \{X(t_{1}),X(t_{2}),\cdots,X(t_{m})\}$ and $\{X(-t_{1}+n),X(-t_{2}+n),\cdots,X(-t_{m}+n)\}$ for every $n$ and $m$ have same joint probabilistic distributions. Under this definition, when $n=t_{1}+t_{m}$, it implies $ \{X(t_{1}),X(t_{2}),\cdots,X(t_{m})\}$ has same joint probability distributions to its symmetric vector $\{X(t_{m}),\cdots,X(t_{2}),X(t_{1})\}$ \cite{Ramsey1995}.

And for a reversible time series, the probability distributions of $X_{m}^{\tau}=\{x_{t},x_{t+\tau},\cdots,x_{t+(m-1)\tau}\}$ for all embedding dimension $m$ and delays $\tau$ should be temporal symmetric \cite{Diks1995}.

These definitions allow us to quantify time irreversibility by measuring the probabilistic divergences of symmetric vectors or the joint probability distributions between the forward and backward time series.

\subsection{Subtraction-based $Y_{S}$ for time irreversibility}
Interestingly, the calculation of joint probability in time irreversibility, involving vector length $m$ and delay $\tau$, has mathematical similarities to the embedding phase space, $X_{m}^{\tau}(i)=\{ x(i),x(i+\tau),\ldots,x(i+(m-1)\tau)\}$, which is also determined by embedding dimension $m$ and delay $\tau$. Inspired by the similarity, a permutation method that maps raw vectors onto order patterns, originally proposed by C. Bandt and B. Pompe \cite{Bandt2002P}, could be a particularly promising solution to simplify the complicate calculation of the joint probability distributions.

\begin{figure}[htb]
  \centering
    \includegraphics[width=7cm,height=4cm]{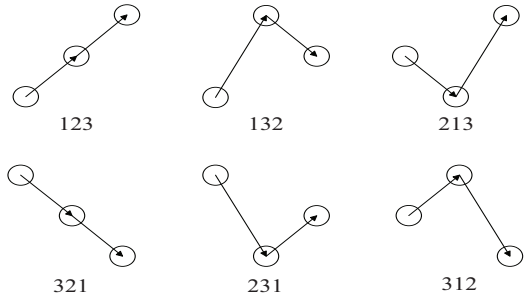}
  \caption{6 order patterns in ascending order of m=3 and $ \tau $=1. The up and down order patterns are symmetric.}
  \label{fig1}
\end{figure}

The permutation method, arising naturally from and inherits causal structures of the time series without any further model assumptions, contains inherent causal dynamical information and the underlying dynamics of temporal structure \cite{Zunino2017P}. It is invariance with respect to nonlinear monotonous transformations and is popular in arbitrary real-world data analysis \cite{Amigo2015,Bandt2016}. Given time series $X(i)$, phase space $X_{m}^{\tau}(i)=\{ x(i),x(i+\tau),\ldots,x(i+(m-1)\tau)\}$ is reconstructed for different embedding dimensions $m$ and delay $\tau$, elements in each vector is reorganized according to their relative values $x_{m\tau}(j_{1}) \leq x_{m\tau}(j_{2}) \leq \cdots \leq x_{m\tau}(j_{i})$, and the indexes of the original values compose the order pattern $\pi_{j}=\{ j_{1},j_{2}, \cdots, j_{i}\}$ whose symmetric permutation is $\pi_{s}=\{j_{i}, \cdots, j_{2},j_{1}\}$. An illustrative example of 6 order patterns in ascending orders of 3-dimensional vectors is depicted in Fig.~\ref{fig1}.

Then we count the amount of each permutation and calculate the probability $p(\pi)$. Time irreversibility or temporal asymmetry could be quantified by measuring the probabilistic difference between $p(\pi_{j})=p(j_{1},j_{2}, \cdots, j_{i})$ and its symmetric $p(\pi_{s})=p(j_{i}, \cdots, j_{2},j_{1})$.

Given the existence of forbidden permutation \cite{Amigo2007,Amigo2010,Carpi2010,Amigo2015,Kulp2017}, some order patterns may not have their symmetric forms. Theoretically, it implies there is significant probabilistic difference if one of the symmetric order patterns does not exist, however case like this leads to the division-based parameters no accountable result. Therefore, we propose a subtraction-based parameter, $Y_{S}$ in Eq.~\ref{eq1} where $p(\pi_{j})\geq p(\pi_{s})$, to measure statistical divergences for time irreversibility.
\begin{eqnarray}
\label{eq1}
Y_{S} = \sum  p(\pi_{j}) \cdot \frac{p(\pi_{j})-p(\pi_{s})}{p(\pi_{j})+p(\pi_{s})}
\end{eqnarray}

The subtraction-based parameter follows some properties for measuring probabilistic difference. Basically, when two probabilities are equal, $Y_{S}=0$. When probabilistic differences between different pairs of symmetric permutations are equal, the larger $p(\pi_{j})/[p(\pi_{j})+p(\pi_{s})]$, the bigger $Y_{S}$ will be. More importantly, if an order pattern do not have its symmetric form, the difference is accountable to be $p(\pi_{j})$.

To quantify the rate of single order patterns without symmetric form, we define Eq.~\ref{eq2}, where $N(\pi)$ is the amount of emergent order types and $N(\pi_{u})$ is the number of single order patterns without symmetric form.
\begin{eqnarray}
\label{eq2}
R_{u}\% = N(\pi_{u})/N(\pi)*100\%
\end{eqnarray}

\subsection{Surrogate data}
Surrogate data is commonly used to identify nonlinearity in time series. The surrogate method is to generate linear surrogate data sets following a null hypothesis that specifies series as some linear process, and detect the nonlinearity by determining whether some statistic aspects of the original and surrogate data are significantly different \cite{Theiler1992}.

Surrogate data could be generated by model-based certain particular autoregressive (AR) process or by model-free methods that randomly adjust amplitude or phase of the original data \cite{Schreiber2000}. In this present paper we adopt the improved amplitude adjusted Fourier transform (iAAFT) \cite{Schreiber1996}, belonging to model-free method, to construct linear data sets that have same autocorrelations, power spectrum and distribution to the data, and we construct a set of 100 surrogate data for each time series. If the discriminating statistic of original time series is smaller than the 2.5th percentile or bigger than 97.5th percentile of the surrogate data set, the null hypothesis is rejected and the time series is said to be time irreversible or nonlinear.

\section{Results}
In this section, nonlinear processes generated by chaotic models, reversible process and their linear surrogate data sets are employed to validate $Y_{S}$, and $Y_{S}$ is applied to identify nonlinearity of epileptic EEG.

\subsection{Time irreversibility in chaotic and reversible series}
Logistic equation, written as $x_{t+1}=r \cdot x_{t}(1-x_{t})$, although simple and deterministic, is an archetypal example to generate dynamical process that is capable of complex and even chaotic behaviors. The bivariate Henon map, given by the coupled equations, $x_{t+1}=1-\alpha \cdot x^{2}_{t}+y_{t}$, $y_{t+1}=\beta \cdot x_{t}$, is an iterated map with chaotic solutions that is popular in the study of dynamical systems. We use the logistic ($r$=4, $x_{1}=0.1$) and Henon ($\alpha$=1.4, $\beta$=0.3 and $x_{1}$=$y_{1}$=0.1) chaotic series and a typical reversible process, Gaussian white noise, to verify our parameter for time irreversibility.

To ensure that all possible permutations will appear, the length of data should not be too short. We select data length from $m!$ to 20$m!$ to determine a proper data length for the three models. And we find that for the logistic and Henon chaotic series, when data length is bigger than 4$m!$, $R_{u}$ has come to be convergent, and as for the Gaussian process, $R_{u}$ becomes to be convergent to be 0 when data length is bigger than 7$m!$. In our paper, we choose a upper bound of $m$=6 and set data length to 10*6!=7200 for the three series.

\begin{table}[htb]
\centering
\caption{$R_{u}\%$ of the logistic, Henon and Gaussian sequences.}
\label{tab1}
\begin{tabular}{ccccc c}
\hline
$m$ &2  &3 &4 &5  &6\\
\hline
logistic 	&0.00 &20.00	&83.33  &93.55  &97.33\\
Henon    	&0.00 &20.00	&53.85  &85.19  &96.55\\
Gaussian    &0.00 &0.00	&0.00  &0.00  &0.00\\
\hline
\end{tabular}
\end{table}

Let us firstly look at the rates of single order patterns, $R_{u}$ listed in tab~\ref{tab1}, of logistic, Henon and Gaussian sequences. For the two nonlinear series, there are extensive occurrence of forbidden permutation that the rates of single order patterns are generally non-zero values and increase with $m$. When $m$ is 3, $R_{u}$ of logistic and Henon series is 20\%, indicating there is one order pattern having no symmetric form, and we find it is the order pattern of '123' in Fig.~\ref{fig1} that its symmetric '321' doesn't exit. When $m$ is 6, the order types of the two chaotic series are almost all unique ones without symmetric versions. As for the Gaussian white noise, $R_{u}$ for $m$ from 2 to 6 are all zero suggesting all permutations have their symmetric forms. The existence of such a large number of single order patterns in chaotic series makes it impossible to use division-based methods to measure the probabilistic differences between the symmetric order patterns.

\begin{figure}[htb]
  \centering
    \includegraphics[width=5cm,height=4cm]{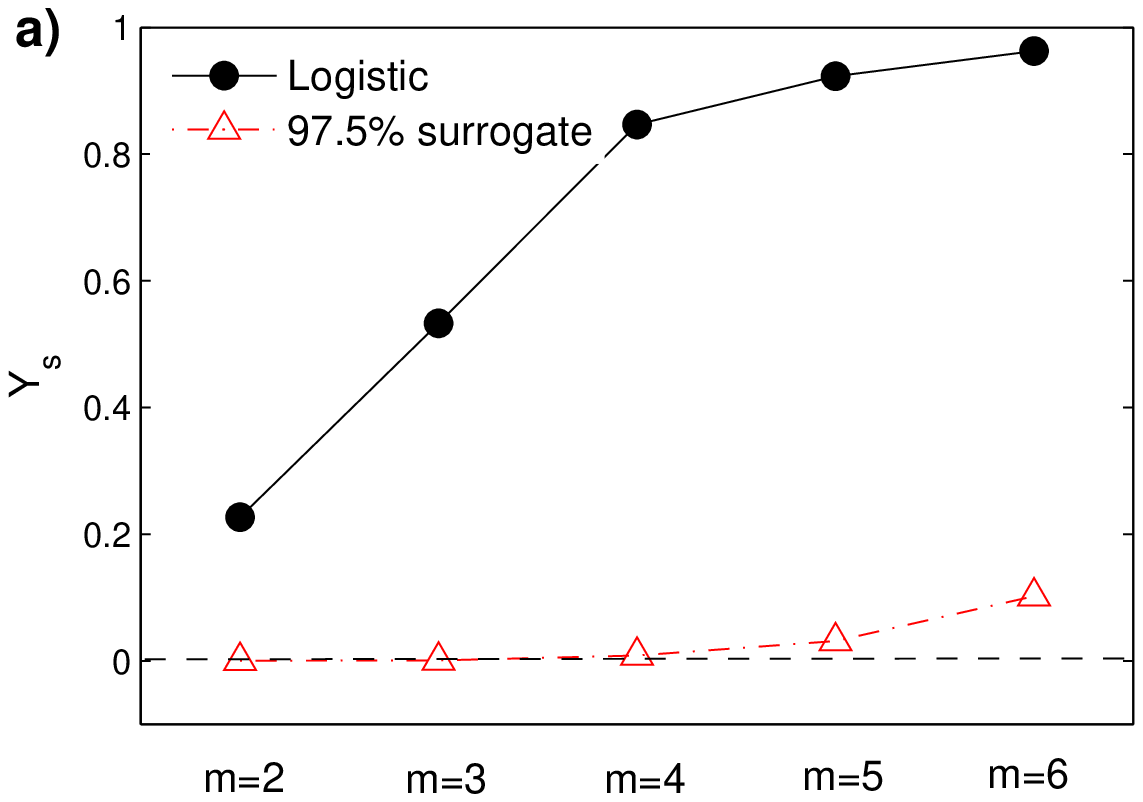}
    \includegraphics[width=5cm,height=4cm]{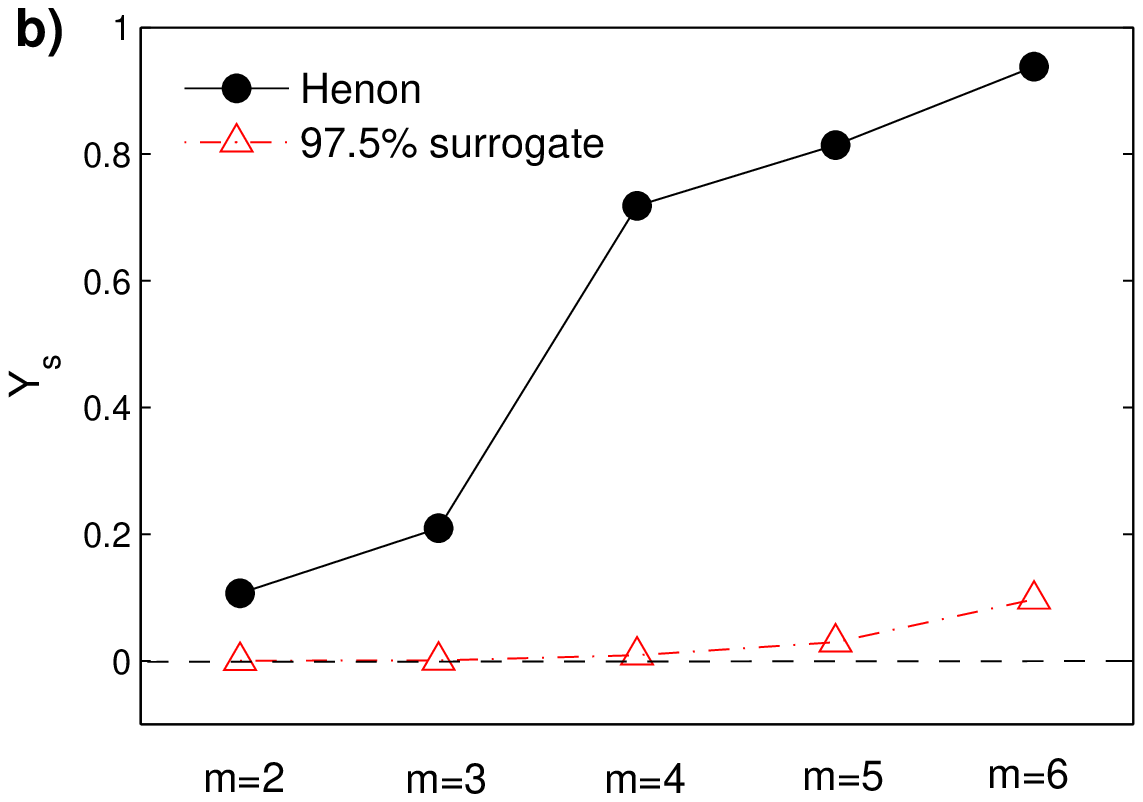}
    \includegraphics[width=5cm,height=4cm]{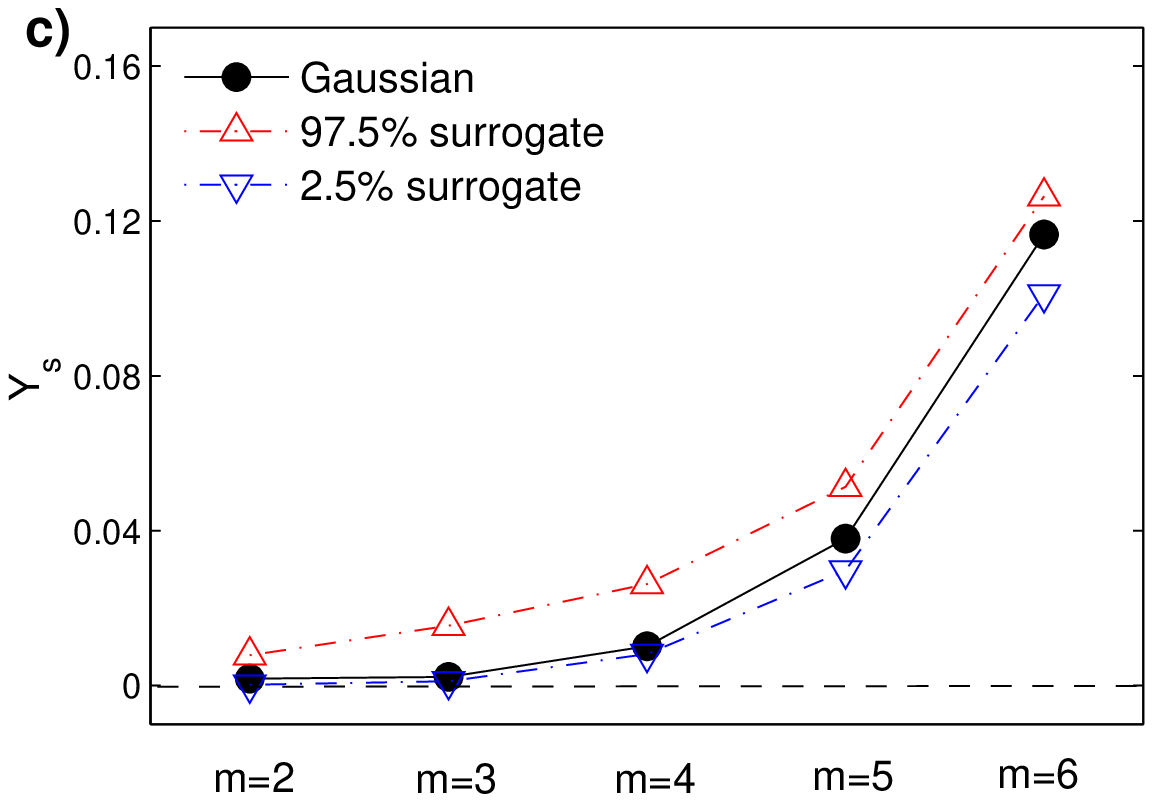}
  \caption{$Y_{S}$ of chaotic and reversible series and the surrogate data. a) Logistic series b) Henon series c) Gaussian process. '97.5\% surrogate' and '2.5\% surrogate' represent the 2.5th and 97.5th percentile of the surrogate data set.}
  \label{fig2}
\end{figure}

Time irreversibility of the three series and their surrogate data are shown in Fig.~\ref{fig2}. $Y_{S}$ of logistic and Henon chaotic series are both larger than the 97.5th percentile of the surrogate data, therefore, from the perspective of surrogate technology, the null hypothesis that the two given series are linear processes is rejected and they are irreversible. However, the null hypothesis should be accepted as for the Gaussian white noise because its $Y_{S}$ are all between the 2.5th and 97.5th percentile of its surrogate data sets. And we find that $Y_{S}$ has similar changes to $R_{u}$ with the increase of $m$, and when $m$ is 6 $Y_{S}$ and $R_{u}$ of logistic and Henon series are all close to 1, indicating close connections between the forbidden permutation and temporal asymmetry.

We test the three processes with different data length from 7000 to 10000, and find that the data length have no significant influence to the results. The two chaotic series and reversible Gaussian process prove that it is effective to quantify the time irreversibility by using $Y_{S}$ to measure the probabilistic differences between symmetric order patterns.

\subsection{Time irreversibility in brain electrical activities}
\begin{figure}[htb]
  \centering
    \includegraphics[width=4cm,height=3.8cm]{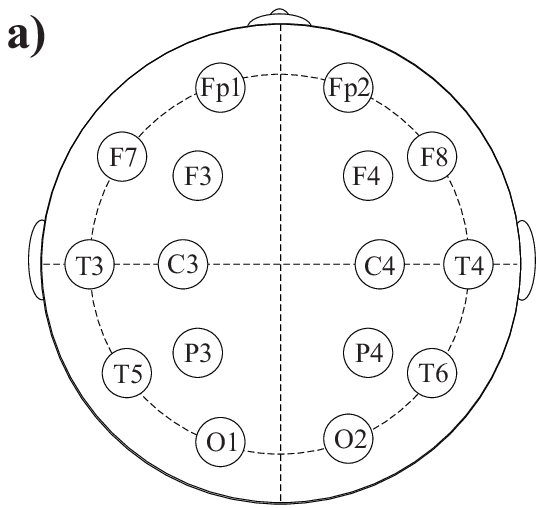}
    \includegraphics[width=5cm,height=4cm]{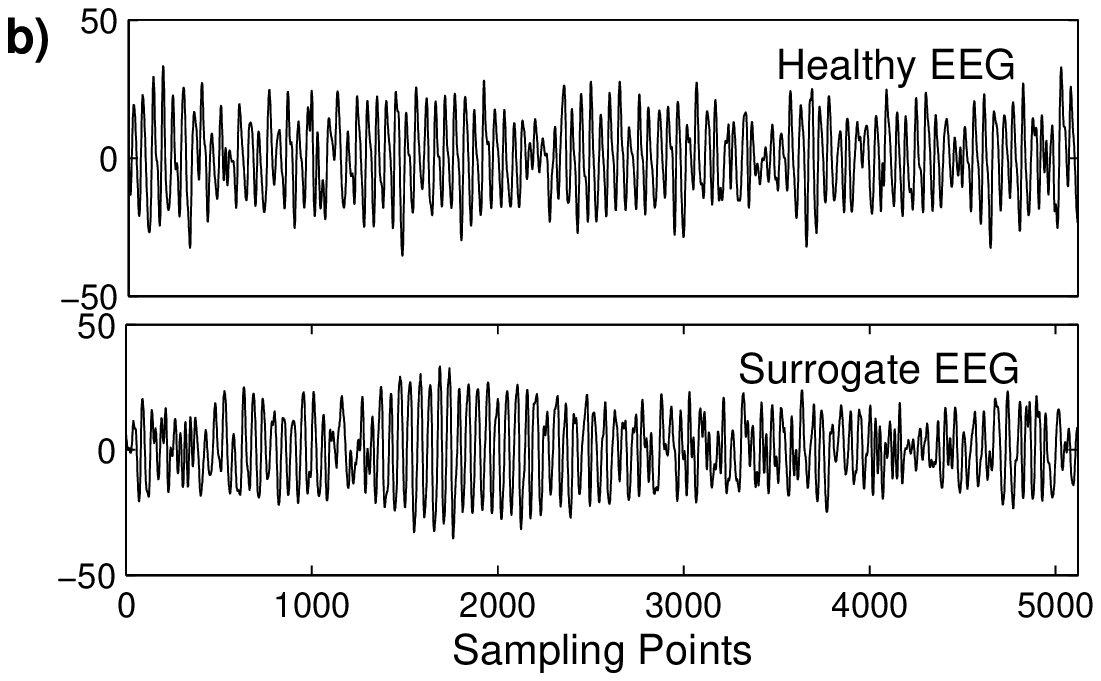}
  \caption{Electrode position of EEG collection and exemplary EEG time series. a) The locations of 16 scalp electrodes. b) Healthy EEG and surrogate data (channel of O2).}
  \label{fig3}
\end{figure}

22 healthy volunteers (aged 15 to 49, mean 26.95$\pm$8.91 years) and 22 epileptic patients (aged 4 to 51, mean 30.0$\pm$13.1 years), enrolled from Nanjing General Hospital of Nanjing Military Command, contribute to multichannel non-intrusive collections of EEGs. Following the standard 10-20 system for electrode position of EEG collection, 8 pairs of symmetric scalp electrodes are located as Fig.~\ref{fig3}a. Eye movement artifacts and noise have been removed. All the subjects contribute EEG collection in duration of about 1 min and with sampling frequency of 512 Hz. Participants are in idle states, and epileptic patients are all in their non-seizure intervals during brain activities monitoring. Exemplary healthy EEG time series and its surrogate data are shown in Fig.~\ref{fig3}b.

Considering the influence of data length on $R_{u}$ and $Y_{S}$, we set $m$=6 and EEG with data length of from 5120 to 30720, in other words with duration of from 10s to 60s, to determine the required minimum data length of the real-world EEG. We find that when data length is bigger than 10240 (20 s), the results come to be convergent, and the results are no affected by the starting points of the EEG data, suggesting $R_{u}$ and $Y_{S}$ are insensitive to the starting recording time.

\begin{figure}[htb]
  \centering
    \includegraphics[width=5cm,height=4cm]{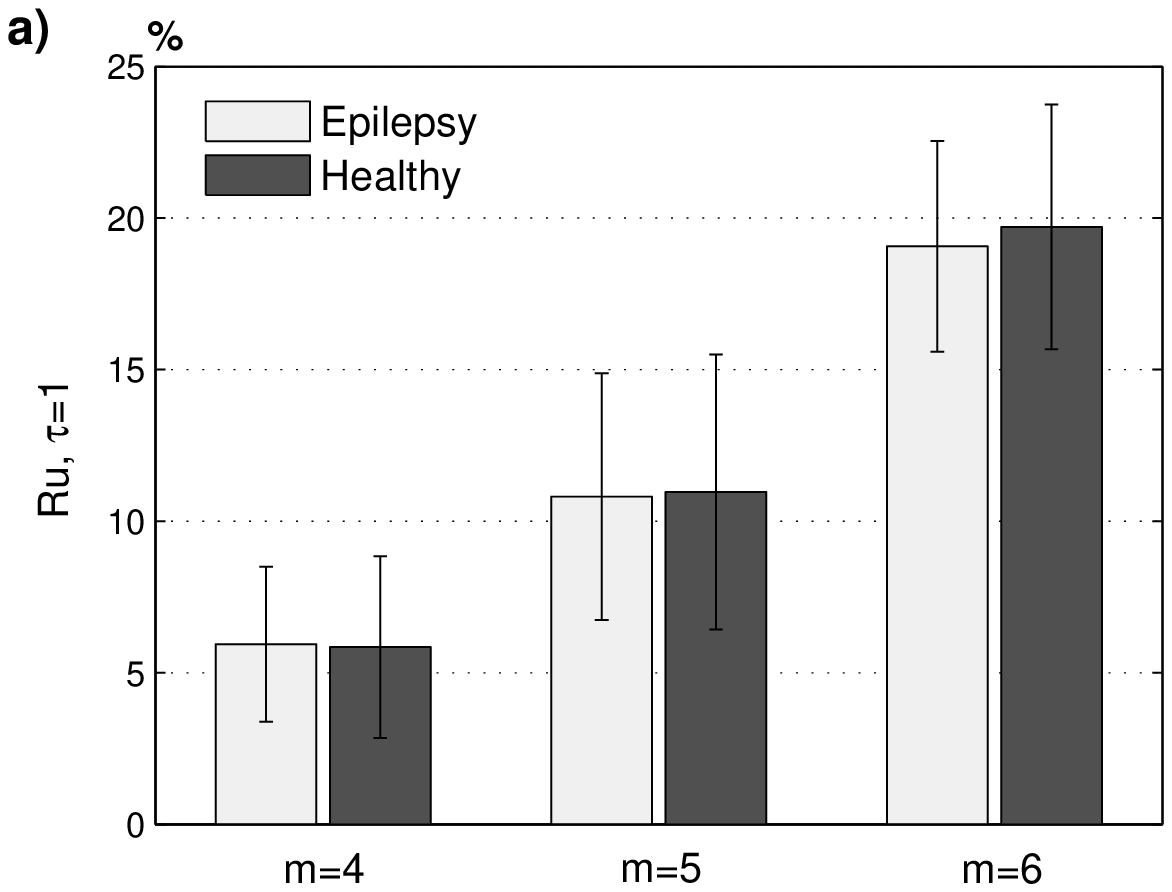}
    \includegraphics[width=5cm,height=4cm]{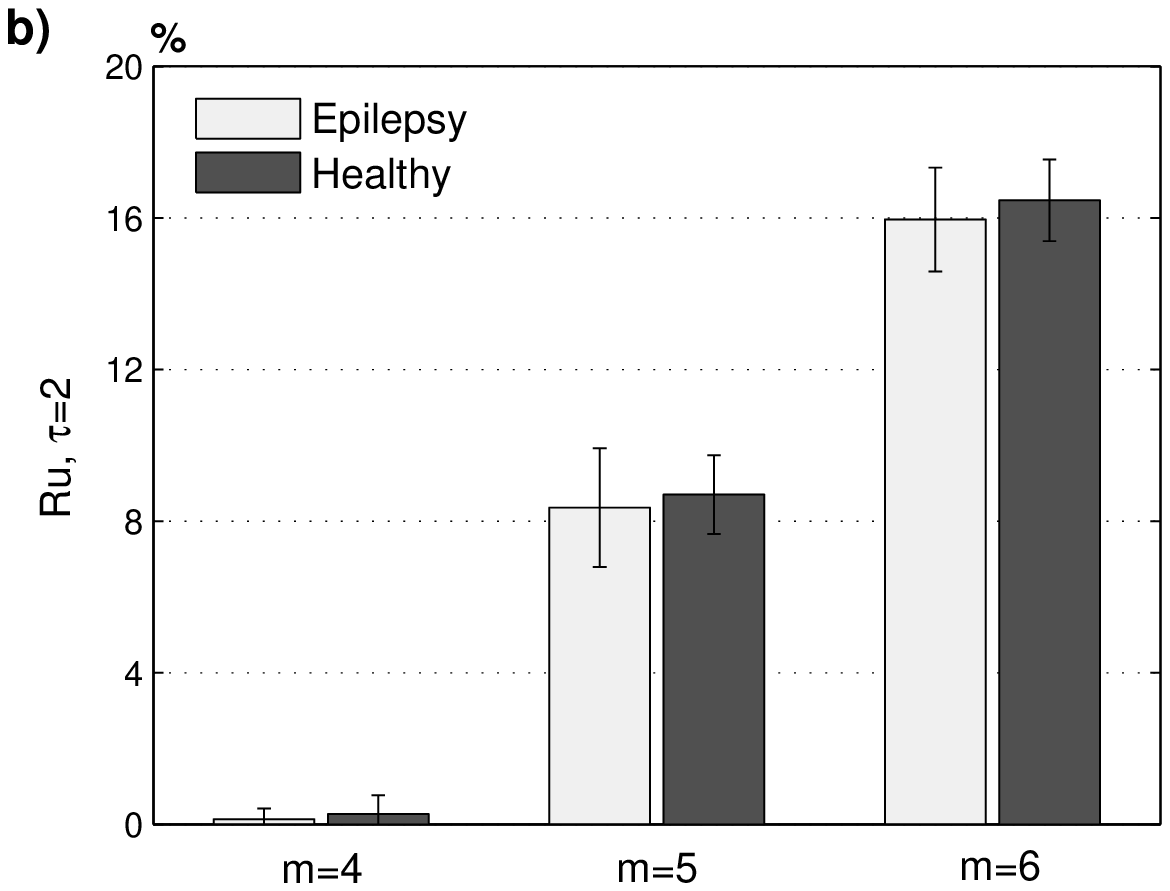}
  \caption{Rates of single order patterns (mean$\pm$std) of the two kinds of EEGs when $\tau$=1 and 2 and $m$=4, 5 and 6. When $m$ are 2 and 3, not displayed in the figures, $R_{u}$=0 that all the order patterns have their symmetric forms.}
  \label{fig4}
\end{figure}

Let us first check the distributions of single order patterns, illustrated in Fig.~\ref{fig4}. When $m$ becomes 4, some motifs loss their symmetries especially when $\tau$=1, and when $m$ becomes 5 and bigger, some more single order types emerge. $R_{u}$ of the healthy EEG are closely equal to and slightly higher than that of the epileptic EEG. From Fig.~\ref{fig4}, we learn that there are also extensive existence of forbidden permutations in the physiological brain data.

\begin{figure}[htb]
  \centering
    \includegraphics[width=5cm,height=4cm]{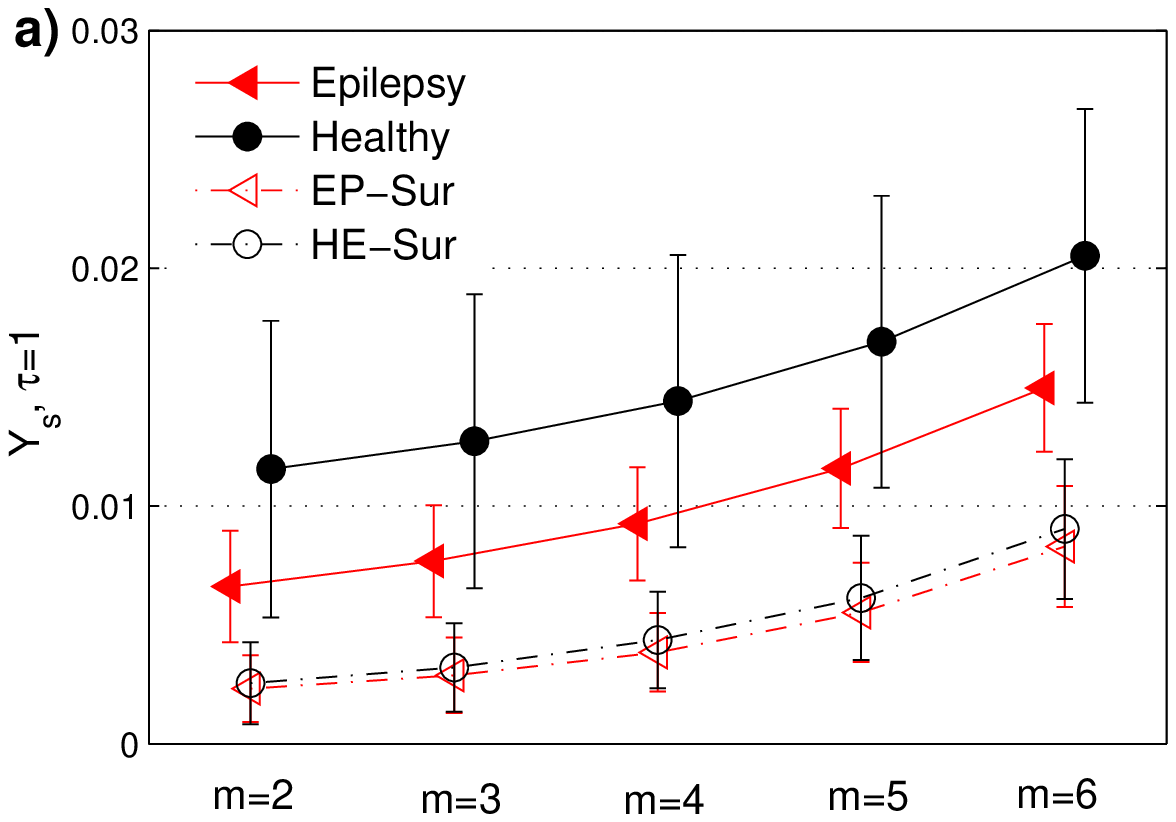}
    \includegraphics[width=5cm,height=4cm]{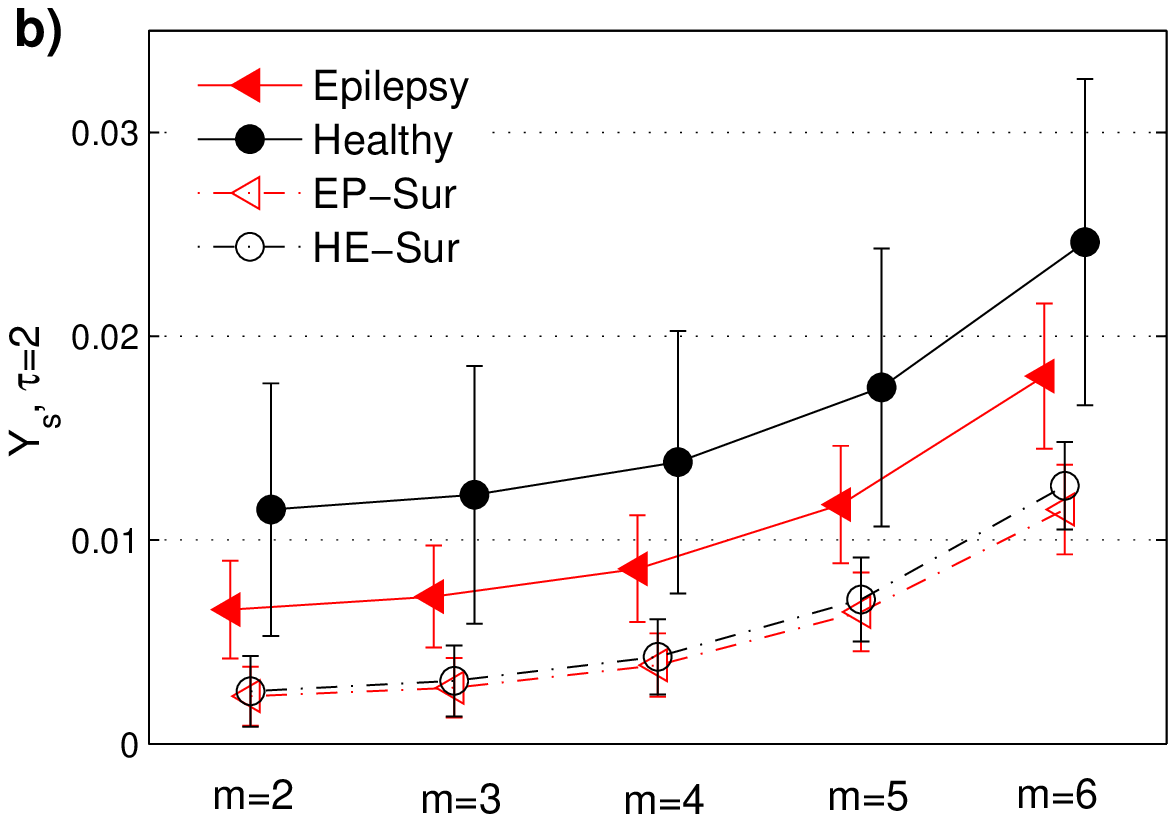}
  \caption{Temporal asymmetry (mean$\pm$std) of the epileptic and healthy subjects. 'EP-Sur' and 'HE-Sur' represent the time irreversibility of epileptic and healthy surrogate.}
  \label{fig5}
\end{figure}

Temporal asymmetry of the healthy and epileptic EEGs and their surrogate data are shown in Fig.~\ref{fig5}, and the results of statistical tests for the discrimination are listed in tab~\ref{tab2}. $Y_{S}$ of both the epileptic and healthy EEGs are higher than the 97.5th percentile of their surrogate data, suggesting both two kinds of brain activities are time irreversible, and time irreversibility of the healthy volunteers are higher than those of the epileptic patients. From tab~\ref{tab2}, the discriminations between the two types of EEGs are acceptable statistically ($p$$<$0.005) and become larger with the increase of length of order patterns or the embedding dimension.

\begin{table}[htb]
\centering
\caption{Independent sample t tests of the temporal asymmetry of the epilepsy and healthy subjects.}
\label{tab2}
\begin{tabular}{ccccc c}
\hline
$\tau \setminus m$ &2  &3 &4 &5  &6\\
\hline
1 	&$2.1*10^{-3}$ &$1.4*10^{-3}$	&$1.1*10^{-3}$  &$9.4*10^{-4}$  &$3.7*10^{-4}$\\
2 	&$2.6*10^{-3}$ &$2.2*10^{-3}$	&$2.2*10^{-3}$  &$2.0*10^{-3}$  &$1.1*10^{-3}$\\
\hline
\end{tabular}
\end{table}

We learn that the parameter quantifying the difference between of symmetric distributions proposed by Costa et al. \cite{Costa2005,Costa2008}, $A = \frac{\sum H(-\Delta x^{-}) - \sum H(\Delta x^{+})}{N (\Delta x \neq 0)}$ where $H$ is the Heaviside function, is equivalent with the simplified $Y_{S}$ of $m$=2 and $\tau$=1, considering the relationship of only two neighboring values, ups $\Delta x^{+}$ and downs $\Delta x^{-}$. $A$ of the epilepsy and healthy EEGs are 0.013$\pm$0.005 and 0.020$\pm$0.008, and the discrimination is statistically acceptable ($p$=$2.1*10^{-3}$), which are in line with our results.

\begin{figure}[htb]
  \centering
    \includegraphics[width=5cm,height=4cm]{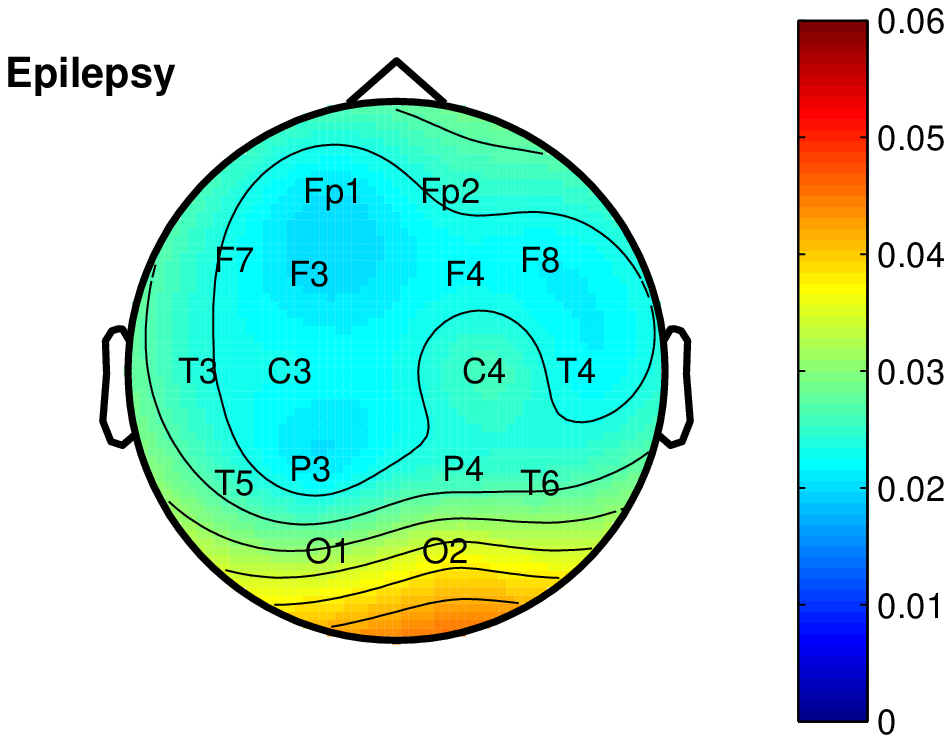}
    \includegraphics[width=5cm,height=4cm]{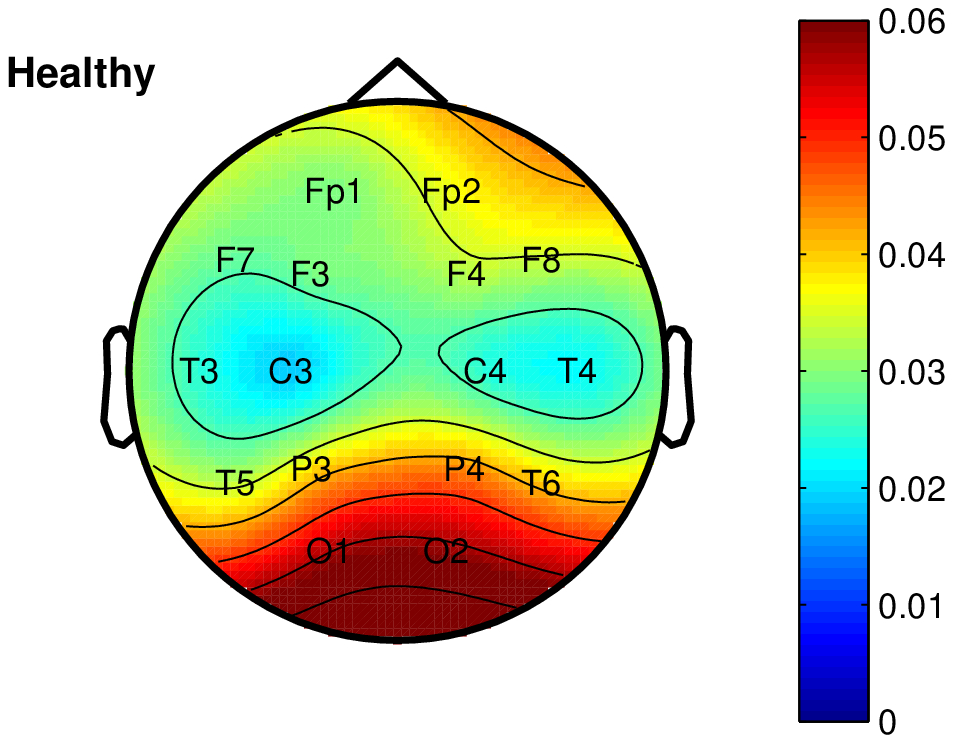}
    \includegraphics[width=6cm,height=4cm]{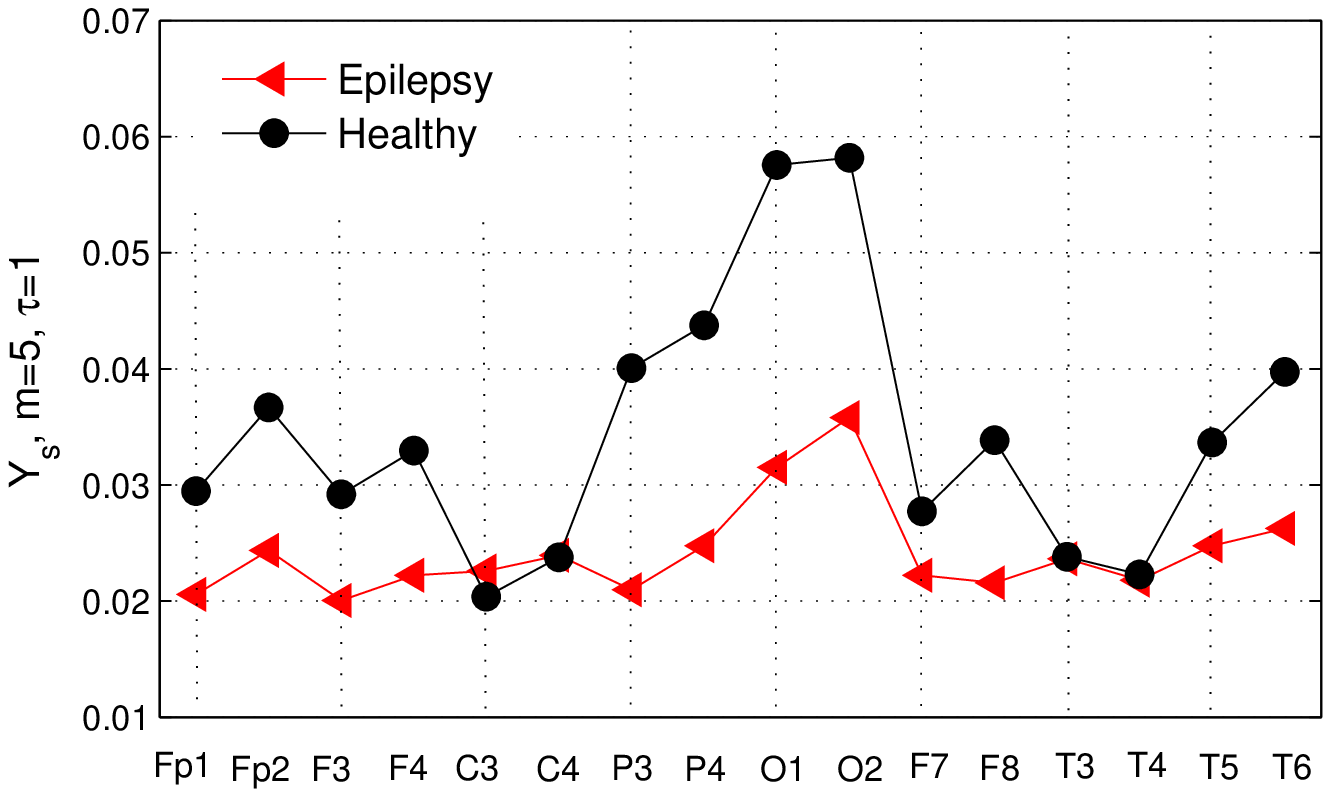}
  \caption{Field maps and line charts of $Y_{S}$ in different brain areas of the epilepsy and healthy.}
  \label{fig6}
\end{figure}

Time irreversibility of different brain areas, taking $m$=5 and $\tau$=1 as an example, are displayed in Fig.~\ref{fig6}, from which we find that the healthy brain electrical activities have overall higher nonlinearity than the seizure-free epileptic ones except the central lobes (C3 and C4). Among other 14 channels, T3 ($p$=0.81) and T4 ($p$=0.74) do not separate the healthy and the epileptic EEGs statistically while other channels all have acceptable discriminations ($p$$<$0.05) between the two kinds of brain behaviors and the parietal lobes (P3, $p$=$1.4*10^{-4}$ and P4, $p$=$6.3*10^{-4}$) have better discriminations.

For one point, both the healthy and the epileptic brain activities manifest the nonlinearity of time irreversibility. Our highly complicated brain, a collection of huge number of neurons interacting and synchronizing with each others, has arguably been a typical complex network \cite{Stam2005,Andrzejak2001} that is subject to various internal physiological factors and external environmental influence, therefore , the brain activities manifest nonlinear dynamics. Brain disorders or diseases, like the all-age affected epilepsy, although have impacts on brain functionality \cite{Litt2002}, the manifestation of inherent nonlinearity of the brain behaviors will not be affected.

For another point, the brain disease of epilepsy cause declination to brain behaviors' nonlinearity. During epileptic seizures, synchronous neuronal firing and severe dysfunction in brain activities could be recorded by EEG \cite{Moshe2015,Iasemidis2003}, while in seizure free intervals the epileptic patients show no significant differences to the normal people. However, according to our report, recurrent seizures bring intriguing changes to brain nonlinear dynamics and lead lasting brain dysfunction. The epileptic patients have significant lower time irreversibility of brain electrical activities than the healthy particularly the brain areas of parietal lobes. The declining nonlinearity in epileptic brain activities might suggest epilepsy causes long-term damage in brain and might lead a step to understanding nonlinear physical characteristics of the brain disease.

\section{Discussions}
From the time irreversibility detections of chaotic and reversible processes and real-world physiological signals, we learn that $R_{u}$, the rate of single order patterns, has close connections with $Y_{S}$ for time irreversibility. $R_{u}$ of logistic and Henon series have similar upward changes with $Y_{S}$ of the two chaotic series, and when $m$$\geq$5, the two parameters are both convergent to 1. As for the reversible Gaussian process, when $m$ is from 2 to 6, $R_{u}$=0 and $Y_{S}$ has no significant difference with the surrogate data. And $R_{u}$ of the healthy EEG are bigger than that of the seizure-free epileptic EEG, which is consistent with the results of $Y_{S}$ although statistical differences of $R_{u}$ ($p$$>$0.6) and $Y_{S}$ ($p$$<$0.005) are different. The connections between $R_{u}$ and $Y_{S}$ suggest that forbidden permutation contain important structural or dynamical information about systems, which is shared by the related reports \cite{Amigo2007,Amigo2010,Carpi2010,Amigo2015,Kulp2017}. And the vector length $m$ and time delay $\tau$ also affect the time irreversibility detection from time series. Logistic and Henon series have convergent $R_{u}$ and $Y_{S}$ when $m$ is bigger than 6. $Y_{S}$ of the healthy and epilepsy EEGs have better statistical discrimination when vector length $m$ increase from 2 to 5 while show no better result when $m$ increase from 5 to 6, and $R_{u}$ and $Y_{S}$ both show no better difference between the two groups of EEGs when $\tau$ increases from 1 to 2. We suppose vector length $m$ and delay $\tau$ may have some connections with the nonlinearity of time series. The dimension $m$ and delay $\tau$ of time irreversibility have mathematical similarity with that of phase space, and they are crucial to attractor reconstruction and nonlinearity detection. According to the embedding theory, if $m$ is smaller than the intrinsic dimension of dynamical systems, the nonlinear dynamical information is still compact and structures cannot be effectively analyzed, and inappropriate delay leads to redundance or irrelevance \cite{Kim1999}. The embedding technique has been paid much attention for its theoretical and applicable significance, and for its interesting similarity to the time irreversibility, physical connections between them and the effects of $m$ and $\tau$ on time irreversibility should gain more deserved attention.

There are some reports \cite{Donges2013,Kulp2017,Andrzejak2001} suggesting epileptic brain activities have higher nonlinearity, which is different from our findings. From the physiological or pathological perspective, the features of epilepsy may lead to the inconsistent results. In epileptic seizures, abnormal firing of neurons and severe nonlinear behaviors \cite{Iasemidis2003} lead extremely higher nonlinearity of brain activities recorded by electrodes on or inside the brain. Moreover the circadian rhythms in human epilepsy \cite{Karoly2018} might lead totally different nonlinear brain behaviors in different stages of epilepsy even all in the seizure-free intervals. Physical reason that may account for the contradictory findings might be the multiple scale theory that these measurements focus on single scale while fail to consider the inherent multiple time scales in healthy brain dynamic activities \cite{Costa2008,Costa2002,Yao2014}. However we would like to emphasize that either physical or physiological explanations should be verified by more representative number of epilepsy patients and more targeted experimental methods, and there might be other possible reasons that contribute to the inconsistent results remaining to be discovered.

When using the ordinal approach for symbolizing time series, equal values, $x_{t}=x_{t+\tau}$, $\tau\neq 0$, might introduce a significant spurious effect, and ignorance of the equalities or broking equalities by adding small random perturbations \cite{Bandt2002P} may lead to false conclusions \cite{Zunino2017P}. This is particularly true when the signal collection is not precision or equal states, for example in the heart rate \cite{Bian2012}, have important information about the systems. Moreover, in Porta or Costa temporal asymmetry index \cite{Costa2005,Costa2008,Porta2008} considering only two adjacent values, equal states imply reversibility, which is not proper when considering more values. In dealing with processes where equal states should not be ignored, one should take equality into serious consideration in applying the permutation method. However, accounting on the continuous distributions of brain electric activities that equal states are very rare, we neglect equal states and consider only inequalities in the EEGs.

\section{Conclusions}
In summary, we simplify the quantification of time irreversibility by transforming m-dimensional vectors into order patterns and measuring probabilistic difference between symmetric order patterns. Considering the existence of forbidden permutation, we proposed a subtraction-based parameter, $Y_{S}$, to measure the probabilistic difference for time irreversibility, which is validated by chaotic and reversible series and surrogate data. The manifestations of time irreversibility of both diseased and healthy EEG is verified, and the declining nonlinearity of seizure-free brain activities might provide valuable information for elucidation of the endogenous mechanisms epileptic brain activities.

\section{Acknowledgment}
The project is supported by the National Natural Science Foundation of China (Grant Nos. 31671006,61771251), Jiangsu Provincial Key R\&D Program (Social Development) (Grant No.BE2015700,BE2016773), Natural Science Research Major Program in Universities of Jiangsu Province (Grant No.16KJA310002).

\nocite{*}

\bibliography{mybibfile}

\end{document}